\documentclass[11pt,a4paper]{article}
\usepackage{a4}
\title{Intensity-intensity correlations as a probe of interferences - under conditions of none in the intensity}

\author{G. S. Agarwal$^{1,2}$, J. von Zanthier$^{1}$, C. Skornia$^{1}$ and H. Walther$^{1}$ \\ 
$^{1}$ Max-Planck-Institut f\"ur Quantenoptik and\\
Sektion Physik der LMU M\"unchen, D-85748 Garching, Germany \\
$^{2}$ Physical Research Laboratory, Navrangpura, Ahmedabad-380 009, India}

\begin{document}
\maketitle

\begin{abstract}

The different behaviour of first order interferences and second order correlations are investigated for the case of two coherently excited atoms. For intensity measurements this problem is equivalent to Young's double slit experiment and was investigated in an experiment by Eichmann et al. [Phys. Rev. Lett. 70, 2359 (1993)] and later analyzed in detail by Itano et al. [Phys. Rev. A 57, 4176 (1998)]. Our results show that in cases where the intensity interferences disappear the intensity-intensity correlations can display an interference pattern with a visibility of up to 100\%. The contrast depends on the polarization selected for the detection and is independent of the strength of the driving field. The nonclassical nature of the calculated intensity-intensity correlations is also discussed.

PACS number(s): 

\end{abstract}

\section{Introduction} \label{Intro}

Young's double slit experiment along with it's modern variants has been central to our understanding of many important aspects of quantum mechanics \cite{feyn,dirac}. In this experiment the interferences arise because the photon can reach the screen either by passing through one or the other slit and it is the inability to distinguish between the two paths that produces the interference fringes. If, however, one could devise a method so as to detect the path the photon took then the interference would be wiped out \cite{zubairy,scully82,scully91,haroche92,walther95,walls95,englert96,wise97}. Young's double slit experiment and other experiments have also been performed with matter waves \cite{Mlynek91,Zeilinger01} where one has clearly understood the disappearance or the fuzziness in the interference pattern if one tries to identify the atomic path, e.g. by detecting the scattered light or probing the internal levels of the diffracted particles \cite{chap95,rempe98,rempe98a,haroche01,pritchard01}. There are also proposals involving cavities to efface the interference by getting {\it welcher weg} information and to recover the interference by using quantum eraser \cite{scully91}. A recent experiment by Bertet et al. \cite{haroche01} follows a scheme very close to the one proposed by Scully and Dr\"uhl. \cite{scully82}. All these experiments provide us with a clear understanding of the close relationship between complementarity, {\it welcher weg} information and the presence or absence of an interference pattern. 

Recently, Eichmann et al. carried out a very interesting experiment where the two slits were replaced by two microscopic objects, namely two Hg$^+$ ions well localised in a linear Paul trap \cite{wine93} (see also \cite{itano98,brewer95,agarwal96,walls97,polder76}). The two ions were driven coherently by a linear polarized laser field close to the $6s ^2 S_{1/2} - 6p ^2 P_{1/2}$ transition in $^{198}$Hg$^+$. To measure the intensity profile of the scattered fluorescence light in the far field they used a polarization selective detection. In this case well defined interference fringes were reported for $\pi$-polarised scattered radiation whereas no interference was found for $\sigma$-polarised emitted light. A detailed theoretical analysis of these findings was given by Itano et al. \cite{itano98}. The results are again interpreted in terms of {\it welcher weg} information: for $\pi$-polarised detection the final states of the two ions are the same as the initial states whereas for $\sigma$-polarised detection they are different and thus one does not (does) have the which path information \cite{itano98}.

In this paper we examine the question whether it is possible to see interference fringes even for $\sigma$-polarised emitted light if one changed the set up and decided to measure other physical quantities. We know from previous work \cite{mollow69} that the radiation emitted by a coherently driven system can have highly nonclassical characteristics. Therefore, in order to understand all the features of the scattered radiation it becomes almost mendatory to study higher order correlations, in particular intensity-intensity correlations of the field emitted by the two atom system \cite{skornia,beige}. In what follows we thus turn our attention to the intensity-intensity second order correlation function. We demonstrate that for two 4-level atoms quantum interferences in the second order correlations of the emitted fluorescence light can be observed for the case of joint detection with two detectors (see Fig. 1). We derive the remarkable result that the depth of modulation in such coincidences can be 100 \% for both $\sigma$ and $\pi$-polarised fluorescence light, independent of the strength of the driving field. This is in strong contrast to the visibility of the interference pattern of the far field {\it intensity} profile which can be observed in case of $\pi$-polarised emitted radiation: this strongly depends on the power of the laser driving the atoms \cite{itano98,polder76}. We can interpret our new outcomes on interferences in the intensity-intensity correlations as resulting from interferences in the two photon decay channels of the system and show that such interferences are absent in case of a single photon decay.

The outline of the paper is as follows: In Section \ref{qintensity} we present a general approach to the problem of interferences in the intensity of emitted fluorescence light and derive basic conditions for the existence of an interference pattern. Here a connection with the traditional {\it welcher weg} - argument is made. In Section \ref{g2} we derive a very general result for the second order intensity-intensity correlation for a two atom system and find conditions for which such a correlation will exhibit interferences. In Section \ref{young} we apply the results of Section \ref{g2} to the experiment of Eichman et al.; here we make new predictions. We use a master equation framework so that we can deal with arbitrarily strong coherent driving fields. The main result of the paper is that in order to observe quantum interferences in the fluorescence light it is not necessarily adequate to study the intensity of the emitted radiation as very interesting interference information can be revealed by studying quantum statistics and in particular the intensity-intensity correlations of the scattered light field.

\section{Conditions for the observation of quantum interferences in the intensity - General considerations} \label{qintensity}

Consider a system of identical atoms located at positions $\vec R_j$. Each atom can in principle involve several emission lines, i.e. emissions at several frequencies. The atom can also produce emission at the same frequency but each emission can come from a different transition. For the purpose of this paper we assume emission at a single frequency coming say from several different transitions $\mid \alpha \ \rangle \rightarrow \mid \beta \ \rangle$. Each atom can either be continuously driven or the initial state might be prepared by a pulsed excitation. 

\vspace{1cm}

{\bf Intensity distribution}

\vspace{0.5cm}

Let $\vec d_{\alpha \beta}$ be the dipole matrix element for the transition $\mid \alpha \ \rangle \rightarrow \mid \beta \ \rangle$. In quantum theory it is known how to relate the statistical properties of the spontaneously emitted radiation to the atomic properties \cite{agarwal74}. In fact, the positive frequency part of the electric field operator in the far field zone can be written in terms of the atomic operators $( \mid \beta \ \rangle \langle \ \alpha \mid )_j$ as

\begin{equation} \label{efield}
\vec E^{(+)} (\vec r , t) = \vec E^{(+)}_0 (\vec r , t) + ( \frac{\omega_0}{c} )^2 \ \frac{e^{i  k r}}{r} \sum_{j, \beta, \alpha} e^{-i k \vec n \cdot \vec R_j} \vec n \times ( \vec n \times \vec d _{\beta \alpha} ) ( \mid \beta \ \rangle \langle \ \alpha \mid )_j 
\end{equation}
\begin{displaymath}
\vec r = \vec n \ r \quad ; \quad k = \frac{\omega_0}{c}
\end{displaymath}

where in Eq. (\ref{efield}) we sum over all transitions corresponding to the possible spontaneous decay channels. Let $\vec \epsilon$ denote the polarization of a detected signal assuming that the detection is polarization selective. The detected signal will then involve the component

\begin{equation} \label{efieldproj}
\vec \epsilon \cdot \vec E^{(+)} (\vec r , t) =  \vec \epsilon \cdot \vec E^{(+)}_0 (\vec r , t) - ( \frac{\omega_0}{c} )^2 \ \frac{e^{i k r}}{r} \sum_{j, \beta, \alpha} e^{-i k \vec n \cdot \vec R_j} ( \vec \epsilon \cdot \vec d _{\beta \alpha} ) ( \mid \beta \ \rangle \langle \ \alpha  \mid )_j
\end{equation}

and hence, except in the forward direction, the intensity of the signal, in arbitrary units, will be 

\begin{equation} \label{intensity}
\begin{array}{ccl}
I & = & \langle \vec \epsilon^{\ast} \cdot \vec E^{(-)} (\vec r, t) \quad \vec \epsilon \cdot \vec E^{(+)} (\vec r, t) \rangle\\
& = & \frac{1}{r^2} ( \frac{\omega_0}{c} )^4 \sum_{j, l, \alpha', \beta', \beta, \alpha} e^{- i k \vec n \cdot (\vec R_j - \vec R_l)} (\vec \epsilon \cdot \vec d_{\beta \alpha} ) (\vec \epsilon^{\ast} \cdot \vec d_{\beta' \alpha'}^{\ast} ) \langle ( \mid \alpha' \ \rangle \langle \ \beta' \mid )_l \ ( \mid \beta \ \rangle \langle \ \alpha \mid )_j \ \rangle
\end{array}
\end{equation}

Let us rewrite Eq.(\ref{intensity}) as

\begin{equation} \label{intensity1}
\begin{array}{ccl}
I & = & \frac{1}{r^2} ( \frac{\omega_0}{c} )^4 \sum_{j, \beta, \alpha, \alpha'} ( \vec \epsilon \cdot \vec d_{\beta \alpha} ) (\vec \epsilon^{\ast} \cdot \vec d_{\beta \alpha'}^{\ast} ) \langle ( \mid \alpha' \ \rangle \langle \ \alpha \mid )_j \rangle + \\
& & 
\frac{1}{r^2} ( \frac{\omega_0}{c} )^4 \sum_{j \not= l, \alpha', \beta', \beta, \alpha} e^{- i k \vec n \cdot (\vec R_j - \vec R_l)} 
( \vec \epsilon \cdot \vec d_{\beta \alpha} ) (\vec \epsilon^{\ast} \cdot \vec d_{\beta' \alpha'}^{\ast} )
\langle ( \mid \alpha' \ \rangle \langle \ \beta' \mid )_l \ ( \mid \beta \ \rangle \langle \ \alpha \mid )_j \ \rangle\\
\end{array}
\end{equation}

As can be seen from (\ref{intensity1}), the intra-atomic interference terms show up in the intensity provided that the correlations 
$\langle ( \mid \alpha' \ \rangle \langle \ \beta' \mid )_l \ ( \mid \beta \ \rangle \langle \ \alpha \mid )_j \ \rangle$ are nonzero.

Of special concern for us is here the case of {\it uncorrelated} atoms. In this case the two atom expectation value factorizes in terms of single atom quantities

\begin{equation} \label{correl2}
\langle ( \mid \alpha' \ \rangle \langle \ \beta' \mid )_l \ ( \mid \beta \ \rangle \langle \ \alpha \mid )_j \ \rangle = \langle ( \mid \alpha' \ \rangle \langle \ \beta' \mid )_l \ \rangle \ \langle ( \mid \beta \ \rangle \langle \ \alpha \mid )_j \ \rangle \quad ; \quad l \not= j
\end{equation}

From (\ref{intensity1}) and (\ref{correl2}) we can see that the radiation from two uncorrelated atoms can interfere only if the atoms have nonzero coherences, i.e. nonzero dipole moments

\begin{equation} \label{correl3}
\langle ( \mid \alpha' \ \rangle \langle \ \beta' \mid )_l  \rangle \not= 0
\end{equation}

This situation is similar to the one occuring in classical electrodynamics, e.g. in the case of interference in the radiation from coherently driven classical antennas. On the other hand, if there are no atomic coherences the interferences can be exhibited only if the atoms are {\it correlated}, as one would expect from (\ref{intensity1}).

\vspace{1cm}

{\bf Nonzero dipole moment and the lack of {\it welcher weg} information}

\vspace{0.5cm}

In quantum theory the existence of interference can be interpreted in terms of {\it welcher weg} information \cite{zubairy}. Thus one would like to understand the interference resulting from nonzero dipole moment as something arising from our lack of information regarding the source of the detected photon. Consider for this purpose for example an initial state of a system of two identical two-level atoms as $\mid e_A , e_B \ \rangle$. The atoms decay independently of each other. There are two paths of decay, i.e. $\mid e_A , e_B \ \rangle \rightarrow \mid g_A , e_B \ \rangle$ or $\mid e_A , e_B \ \rangle \rightarrow \mid e_A , g_B \ \rangle$. For these two paths the final state of the two atom system is different and therefore interference does not occur. Let us next consider as initial state a state which is a superposition of ground and excited levels $\mid \psi \ \rangle = c_e \mid e \ \rangle + c_g \mid g \ \rangle$. Note that this is a state for which the dipole moment is nonzero. Let the initial state of the two atom system be $\mid \psi_A , \psi_B \ \rangle$. After decay of a photon we get depending on which path of decay is considered,

\begin{equation} \label{decay1}
\mid \psi_A , \psi_B \ \rangle \rightarrow c_e \mid g_A \ \rangle \mid \psi_B \ \rangle
\end{equation}
or
\begin{equation}
\mid \psi_A , \psi_B \ \rangle \rightarrow c_e \mid \psi_A \ \rangle \mid g_B \ \rangle
\end{equation}

The two paths lead to final states which have a common component $\mid g_A g_B \rangle$ occuring with amplitude  $c_e c_g$. Thus an interference appears which is proportional to $\mid c_e c_g \mid^2$ or to the modulus square of the dipole moment $c^{\ast}_e c_g$. In this manner we have established a connection between arguments based on {\it welcher weg} information and the
existence of a dipole moment.

\section{Quantum interferences in intensity-intensity correlations - General result for {\it uncorrelated} atoms}   \label{g2}

From our discussion in Section \ref{qintensity} it is clear that the case of {\it uncorrelated} atoms with zero dipole moment is especially challenging since in this case no quantum interferences are observable in the intensity profile of the far field, i.e. $ \langle I \rangle$ shows no modulation. In this Section we demonstrate on very general grounds that nevertheless quantum interferences with high modulation depth may exist, in particular if the intensity-intensity correlations in the field produced by the uncorrelated atoms are considered.

In order to motivate our discussion consider again the simple case of two identical 2-level atoms with initial state $\mid e_A , e_B \ \rangle$. Consider the following two-photon emission channels (see Fig. 1)

\begin{equation} \label{decay2}
\mid e_A , e_B \ \rangle \rightarrow \mid g_A \ \rangle \mid e_B \ \rangle \mid \vec k_i \ \rangle \rightarrow \mid g_A \ \rangle \mid g_B \ \rangle \mid \vec k_1 \vec k_2 \ \rangle
\end{equation}
or
\begin{equation}
\mid e_A , e_B \ \rangle \rightarrow \mid e_A \ \rangle \mid g_B \ \rangle \mid \vec k_i \ \rangle \rightarrow \mid g_A \ \rangle \mid g_B \ \rangle \mid \vec k_1 \vec k_2 \ \rangle \quad ,
\end{equation}
\begin{displaymath}
i = 1 , 2
\end{displaymath}

Clearly, the different paths for two photon decay can interfere as the initial and final states of the paths are identical. This can be demonstrated explicitly by using higher order Fermi-Golden rule. The phase factors like $e^{i \vec k_{\alpha} \cdot \vec R_{\beta}}$ originating from the intermediate states give rise to the interference terms.

We now consider the general situation for a two atom system using the result (\ref{efieldproj}). Let us write it in the form

\begin{equation} \label{efieldproj1}
\vec \epsilon_1 \cdot \vec E^{(+)} (\vec r_1 , t) =  \vec \epsilon_1 \cdot \vec E^{(+)}_0 (\vec r_1 , t) + \mathcal{E}^{(+)}_A ( 1 ) + \mathcal{E}^{(+)}_B ( 1 ) 
\end{equation}

where for example $\mathcal{E}^{(+)}_A ( 1 )$ is the contribution to the scattered field by atom $A$ at the point $\vec r_1$, with polarization $\vec \epsilon_1$. In what follows we examine the second order intensity-intensity correlation for measurements with two detectors located at $\vec r_1$ and $\vec r_2$. We furthermore suppose a polarization selective detection, e.g. we assume that the detector at $\vec r_{\alpha}$ selects polarization $\vec \epsilon_{\alpha}$. Let us consider the intensity-intensity correlation function defined by

\begin{equation} \label{G2a}
\begin{array}{ccl}
G^{(2)} (  \vec r_1 , \vec \epsilon_1 , t ; \vec r_2, \vec \epsilon_2 , t ) & = & G^{(2)} ( 1 , 2 )\\
& = & \langle \vec \epsilon^{\ast}_1 \cdot \vec E^{(-)} ( \vec r_1 , t ) \ \vec \epsilon^{\ast}_2 \cdot \vec E^{(-)} ( \vec r_2 , t ) \ \vec \epsilon_2 \cdot \vec E^{(+)} ( \vec r_2 , t ) \ \vec \epsilon_1 \cdot \vec E^{(+)} ( \vec r_1 , t ) \ \rangle
\end{array}
\end{equation}

which on using (\ref{efieldproj1}) reduces to

\begin{equation} \label{g2simple}
\begin{array}{ccl}
G^{(2)} ( 1 , 2 ) & = & \langle ( \mathcal{E}^{(-)}_A ( 1 ) + \mathcal{E}^{(-)}_B ( 1 ) ) \ ( \mathcal{E}^{(-)}_A ( 2 ) + \mathcal{E}^{(-)}_B ( 2 ) ) \\
& & ( \mathcal{E}^{(+)}_A ( 2 ) + \mathcal{E}^{(+)}_B ( 2 ) ) \ ( \mathcal{E}^{(+)}_A ( 1 ) + \mathcal{E}^{(+)}_B ( 1 ) ) \ \rangle
\end{array}
\end{equation}

Note that the vacuum terms do not contribute to normally ordered correlations. Since the single atom operators satisfy the property

\begin{equation}
\mid \alpha \ \rangle \langle \beta \mid \alpha' \ \rangle \langle \beta' \mid = \delta_{\beta \alpha'} \mid \alpha \ \rangle \langle \beta' \mid
\end{equation}

the terms like $\mathcal{E}^{(-)}_A ( 1 ) \mathcal{E}^{(-)}_A ( 2 )$ are identically zero. Hence (\ref{g2simple}) reduces further to 

\begin{equation}  \label{g2simplea}
\begin{array}{ccl}
G^{(2)} ( 1 , 2 ) & = & \langle ( \mathcal{E}^{(-)}_A ( 1 ) \mathcal{E}^{(-)}_B ( 2 ) +  \mathcal{E}^{(-)}_B ( 1 ) \mathcal{E}^{(-)}_A ( 2 ) ) \\
& & ( \mathcal{E}^{(+)}_B ( 2 ) \mathcal{E}^{(+)}_A ( 1 ) + \mathcal{E}^{(+)}_A ( 2 ) \mathcal{E}^{(+)}_B ( 1 ) ) \ \rangle
\end{array}
\end{equation}

We next make use of the {\it uncorrelated} nature of the atoms $A$ and $B$ to simplify (\ref{g2simplea}) in the following manner

\begin{equation} \label{g2simpleb}
\begin{array}{ccl}
G^{(2)} ( 1 , 2 ) & = &
\langle \mathcal{E}^{(-)}_A ( 1 ) \mathcal{E}^{(+)}_A ( 1 ) \ \rangle \langle \mathcal{E}^{(-)}_B ( 2 ) \mathcal{E}^{(+)}_B ( 2 ) \ \rangle\\
& &  + \langle \mathcal{E}^{(-)}_A ( 1 ) \mathcal{E}^{(+)}_A ( 2 ) \ \rangle \langle \mathcal{E}^{(-)}_B ( 2 ) \mathcal{E}^{(+)}_B( 1 ) \ \rangle\\
& &  + \langle \mathcal{E}^{(-)}_A ( 2 ) \mathcal{E}^{(+)}_A ( 1 ) \ \rangle \langle \mathcal{E}^{(-)}_B ( 1 ) \mathcal{E}^{(+)}_B ( 2 ) \ \rangle\\
& &  + \langle \mathcal{E}^{(-)}_A ( 2 ) \mathcal{E}^{(+)}_A ( 2 ) \ \rangle \langle \mathcal{E}^{(-)}_B ( 1 ) \mathcal{E}^{(+)}_B ( 1 ) \ \rangle
\end{array}
\end{equation}

Clearly, the existence of interference terms in $G^{(2)} ( 1 , 2 )$ depends on the nonvanishing of the {\it amplitude correlation} function

\begin{equation} \label{g1}
G^{(1)}_A ( 1 , 2 ) = \langle \mathcal{E}^{(-)}_A ( 1 ) \mathcal{E}^{(+)}_A ( 2 ) \rangle
\end{equation}

Note that $G^{(1)}_A ( 1 , 2 )$, which is a measure of spatial coherence, is not
necessarily zero even if $\langle \mathcal{E}^{(-)}_A ( 1 ) \rangle$ is zero.

We can rewrite (\ref{g2simpleb}) also in the form

\begin{equation} \label{g2ssimple}
G^{(2)} ( 1 , 2 ) = ( G^{(1)}_A ( 1 , 1 ) G^{(1)}_B ( 2 , 2 ) + G^{(1)}_A ( 2 , 2 ) G^{(1)}_B ( 1 , 1 ) ) \cdot \lbrack 1 + \Gamma^{(2)} ( 1 , 2 ) \rbrack \quad ,
\end{equation}

where

\begin{equation} \label{Gamma}
\Gamma^{(2)} ( 1 , 2 ) = \frac{( G^{(1)}_A ( 1 , 2 ) G^{(1)}_B ( 2 , 1 ) + c. c. )}{\lbrace G^{(1)}_A ( 1 , 1 ) G^{(1)}_B ( 2 , 2 ) + G^{(1)}_A ( 2 , 2 ) G^{(1)}_B ( 1 , 1 ) \rbrace}
\end{equation}

Note that (\ref{g2ssimple}) has resemblence to the well known result for thermal light \cite{zubairy}. However, it should be borne in mind that for radiation produced by coherently driven single atoms $\Gamma^{(2)} ( 1 , 2 )$ can also be negative. 

Let us now examine more closely the structure of (\ref{g1}). By using (\ref{efieldproj}), we obtain

\begin{equation} \label{G1mod}
\begin{array}{ccl}
G^{(1)}_A ( 1 , 2 ) & = &
( \frac{\omega_0}{c} )^4 \frac{e^{i k (r_2 - r_1)}}{r_1 r_2} e^{- i k ( \vec n_2 - \vec n_1 ) \cdot \vec R_A}
\sum  (\vec \epsilon_2 \cdot \vec d_{\beta \alpha} ) (\vec \epsilon^{\ast}_1 \cdot \vec d_{\beta' \alpha'}^{\ast} ) \langle ( \mid \alpha' \ \rangle \langle \ \beta' \mid \beta \ \rangle \langle \ \alpha \mid  \ \rangle\\
& = &
( \frac{\omega_0}{c} )^4 \frac{e^{i k (r_2 - r_1)}}{r_1 r_2} e^{- i k ( \vec n_2 - \vec n_1 ) \cdot \vec R_A}
\sum  (\vec \epsilon_2 \cdot \vec d_{\beta \alpha} ) (\vec \epsilon^{\ast}_1 \cdot \vec d_{\beta \alpha'}^{\ast} ) \rho_{\alpha \alpha'}
\end{array}
\end{equation}

which becomes 

\begin{equation} \label{G1mod1}
\begin{array}{ccl}
G^{(1)}_A ( 1 , 2 ) & = &
( \frac{\omega_0}{c} )^4 \frac{e^{i k (r_2 - r_1)}}{r_1 r_2} e^{- i k ( \vec n_2 - \vec n_1 ) \cdot \vec R_A}
\sum  (\vec \epsilon_2 \cdot \vec d_{\beta \alpha} ) (\vec \epsilon^{\ast}_1 \cdot \vec d_{\beta \alpha}^{\ast} ) \rho_{\alpha \alpha} \quad ,
\end{array}
\end{equation}

if there are no excited state coherences, i.e. if $\rho_{\alpha \alpha'} = 0$ \cite{footnot}. In this case $G^{(1)}_A ( 1 , 2 )$ is nonvanishing as long as $\vec \epsilon_2 \cdot \vec \epsilon_1^{\ast} \not= 0$.

For a two level transition $\mid \alpha \ \rangle \leftrightarrow \mid  \beta \ \rangle$ where $\mid \alpha \ \rangle$ ($\mid \beta \ \rangle$) represents the excited (ground) state (\ref{G1mod}) can be simplified to

\begin{equation} \label{G1moda}
G^{(1)}_A ( 1 , 2 ) = 
( \frac{\omega_0}{c} )^4 \frac{e^{i k (r_2 - r_1)}}{r_1 r_2} e^{- i k ( \vec n_2 - \vec n_1 ) \cdot \vec R_A}
(\vec \epsilon_2 \cdot \vec d_{\beta \alpha} ) (\vec \epsilon^{\ast}_1 \cdot \vec d_{\beta \alpha}^{\ast} ) \rho_{\alpha \alpha}
\end{equation}

On substituting in (\ref{Gamma}) we get

\begin{equation}
\Gamma^{(2)} ( 1 , 2 ) = \cos ( \ k \ (\vec n_2 - \vec n_1 ) \cdot ( \vec R_A - \vec R_B ) )
\end{equation}

The interference pattern of the intensity-intensity second order correlation function $G^{(2)} ( 1 , 2 )$ can thus exhibit a modulation depth of 100 \% irrespective of the strength of the driving field, i.e. irrespective of the degree of excitation of the atom. In particular, for the initial state $\mid e_A , e_B \ \rangle$ in absence of a continuous coherent drive, one obtains well defined interferences in the second order correlation function $G^{(2)} ( 1 , 2 )$ whereas no interference fringes can be seen in the far field intensity distribution $\langle I \rangle$ as already shown in Section \ref{qintensity}.

\section{Young's interference experiment by Eichmann et al. revisited - new predictions for cases when no interferences were observed in the intensity}   \label{young}

In this Section we will reexamine the experiment by Eichmann et al. introduced above in Section \ref{Intro} \cite{wine93,itano98}. For that purpose we make use of a master equation approach for the atomic dynamics which has the advantage to be able to deal with arbitrarily strong coherent driving fields. Thus unlike in the work of Itano et al. \cite{itano98} we do not use a perturbation theoretical approach. In what follows we demonstrate how quantum interferences can be recovered in the intensity-intensity correlations even in case of $\sigma$-polarized light. In this case no interferences were observed in the far field intensity distribution \cite{wine93,itano98}.

The corresponding level scheme is shown in Fig. 2. Eichmann et al. considered excitation of two 4-level atoms by $\pi$-polarized light \cite{wine93}. The spontaneous emission could occur on the transitions $\mid 1 \ \rangle \rightarrow \mid 2 \ \rangle$ ; $\mid 3 \ \rangle \rightarrow \mid 4 \ \rangle$ ($\pi$-polarization) and on the transitions $\mid 1 \ \rangle \rightarrow \mid 4 \ \rangle$ ; $\mid 3 \ \rangle \rightarrow \mid 2 \ \rangle$ ($\sigma$-polarization) (see Fig. 2). Assuming excitation by resonant light, the density matrix equations for this system can be written in the form

\begin{equation} \label{mastereq}
\begin{array}{ccl}
\dot \rho_{11} & = & i g \rho_{21} - i g \rho_{12} - 2 ( \gamma_0 + \gamma ) \rho_{11}\\
\dot \rho_{12} & = & i g \rho_{22} - i g \rho_{11} - ( \gamma_0 + \gamma ) \rho_{12}\\
\dot \rho_{13} & = & i g \rho_{23} + i g \rho_{14} - 2 ( \gamma_0 + \gamma ) \rho_{13}\\
\dot \rho_{14} & = & i g \rho_{24} + i g \rho_{13} - ( \gamma_0 + \gamma ) \rho_{14}\\
\dot \rho_{22} & = & - i g \rho_{21} + i g \rho_{12} + 2 \gamma_0 \rho_{11} + 2 \gamma \rho_{33}\\
\dot \rho_{23} & = & i g \rho_{13} + i g \rho_{24} - ( \gamma_0 + \gamma ) \rho_{23}\\
\dot \rho_{24} & = & i g \rho_{14} + i g \rho_{23}\\
\dot \rho_{33} & = & - i g \rho_{43} + i g \rho_{34} - 2 ( \gamma_0 + \gamma ) \rho_{33}\\
\dot \rho_{34} & = & - i g \rho_{44} - i g \rho_{33}\\
\end{array}
\end{equation}

The remaining equations can be generated by taking complex conjugates or using $\textrm{Tr} \lbrace \rho \rbrace = 1$. Here $2 g$ is the Rabi frequency of the driving field $g = ( \vec d_{12} \cdot \vec \epsilon / \hbar )$ and $2 \gamma_0$ and $2 \gamma$ are the rates of spontaneous emission as shown in Fig. 2. Using (\ref{mastereq}), the steady state solutions are found to be

\begin{equation} \label{masterss1}
\rho_{13} = \rho_{23} = \rho_{24} = \rho_{14} = 0
\end{equation}

\begin{equation} \label{masterss2}
\begin{array}{ccccl}
\rho_{11} & = & \rho_{33} & = & \frac{1}{2} \frac{g^2}{( 2 g^2 + ( \gamma_0 + \gamma )^2 )}\\
\rho_{22} & = & \rho_{44} & = & 1 - \frac{1}{2} \frac{( 3 g^2 - ( \gamma_0 + \gamma )^2 )}{( 2 g^2 + ( \gamma_0 + \gamma )^2 )}\\
\rho_{12} & = & - \rho_{34} & = & \frac{i}{2} \frac{g \ ( \gamma_0 + \gamma )}{( 2 g^2 + ( \gamma_0 + \gamma )^2 )}\\
\end{array}
\end{equation}

For the present system we therefore get, using (\ref{efieldproj}) and Fig. 2

\begin{equation} \label{Xi1}
\begin{array}{ccl}
\mathcal{E}^{(+)}_A ( 1 ) & = & - ( \frac{\omega_0}{c} )^2 \frac{e^{i k r_1}}{r_1}
e^{- i k (\vec n_1 - \vec n_l) \cdot \vec R_A} \\
& & \lbrace \vec \epsilon_1 \cdot \vec d_{21} \mid 2 \ \rangle \langle 1 \mid + \vec \epsilon_1 \cdot \vec d_{43} \mid 4 \ \rangle \langle 3 \mid \\
& & +  \vec \epsilon_1 \cdot \vec d_{41} \mid 4 \ \rangle \langle 1 \mid +  \vec \epsilon_1 \cdot \vec d_{23} \mid 2 \ \rangle \langle 3 \mid \rbrace \quad ,
\end{array}
\end{equation}

where $\vec n_l$ is the direction of the exciting radiation. Note that $\vec d_{21} \| \ \vec d_{43}$, $\vec d_{41} \ \bot \ \vec d_{23}^{\ast}$, $\vec d_{41} \ \bot \ \vec d_{21}$ etc. Thus, if the polarization vector $\vec \epsilon_1$ is chosen such that $\vec \epsilon_1 \ \bot \ \vec d_{21}$, then one obtains from (\ref{masterss1}) and (\ref{Xi1})

\begin{equation} \label{Xi2}
\begin{array}{ccl}
\langle \mathcal{E}_A^{(+)} ( 1 ) \rangle & = & - ( \frac{\omega_0}{c} )^2 \frac{e^{i k r_1}}{r_1} e^{- i k (\vec n_1 - \vec n_l) \cdot \vec R_A} \lbrace \vec \epsilon_1 \cdot \vec d_{41} \rho_{14} +  \vec \epsilon_1 \cdot \vec d_{23} \rho_{32} \rbrace = 0
\end{array} 
\end{equation}
\begin{displaymath}
\textrm{if} \quad \vec \epsilon_1 \cdot \vec d_{21} = 0 
\end{displaymath}

On the other hand, for $\vec \epsilon_1 \ \bot \ \vec d_{41}$ one gets

\begin{equation} \label{Xi3}
\begin{array}{ccl}
\langle \mathcal{E}_A^{(+)} ( 1 ) \rangle & = & - 2 \ ( \frac{\omega_0}{c} )^2 \frac{e^{i k r_1}}{r_1} e^{- i k (\vec n_1 - \vec n_l) \cdot \vec R_A} \ ( \vec \epsilon_1 \cdot \vec d_{21} ) \ \rho_{12} \not= 0
\end{array}
\end{equation}
\begin{displaymath}
\textrm{if} \quad \vec \epsilon_1 \cdot \vec d_{41} = 0 
\end{displaymath}

Thus the mean dipole moment or the mean radiated field in case of $\pi$-polarized emitted light is nonzero (\ref{Xi3}) whereas it vanishes for $\sigma$-polarization (\ref{Xi2}). Therefore, as shown in Section \ref{qintensity}, interference fringes can be observed only for $\pi$-polarized fluorescence but not for $\sigma$-polarized fluorescence light. This is in agreement with the experimental and theoretical results of Wineland's group \cite{wine93,itano98}. To show this more explicitly we present in what follows the general result for the far field intensity distribution of the emitted fluorescence light.

\vspace{1cm}

{\bf Far field intensity pattern for arbitrary polarisation}

\vspace{0.5cm}

Using (\ref{Xi1}) and the solutions (\ref{masterss1}) and (\ref{masterss2}) we find

\begin{equation} \label{G11}
G^{(1)}_A ( 1 , 1 ) = ( \frac{\omega_0}{c} )^4 \frac{1}{r_{1}^{2}} \ \rho_{11} \cdot \lbrace
\mid \vec \epsilon_1 \cdot \vec d_{21} \mid^2 + 
\mid \vec \epsilon_1 \cdot \vec d_{43} \mid^2 + 
\mid \vec \epsilon_1 \cdot \vec d_{41} \mid^2 + 
\mid \vec \epsilon_1 \cdot \vec d_{23} \mid^2 \rbrace
\end{equation}

where the different dipole matrix elements can be shown to be given by \cite{sobelman}

\begin{equation} \label{dipole}
\begin{array}{cccll}
\vec d_{21} & = & - \vec d_{43} & = & - \mathcal{D} \ \hat z\\
\vec d_{41} & = & \vec d_{23}^{\ast} & = & \frac{\mathcal{D}}{\sqrt{3}} \ \hat \epsilon_{-} \quad , \\
\end{array}
\end{equation}

with $\hat \epsilon_{-} = \frac{\hat x - i \hat y}{\sqrt{2}}$. In (\ref{dipole}) $\mathcal{D}$ denotes the reduced matrix element of the dipole operator $\vec d$. On substituting (\ref{dipole}) into (\ref{G11}) we obtain

\begin{equation} \label{G11a}
G^{(1)}_A ( 1 , 1 ) = ( \frac{\omega_0}{c} )^4 \frac{1}{r_{1}^{2}} \ \rho_{11} \ (\frac{\mathcal{D}^2}{3}) \ , 
\end{equation}

which also turns out to be equal to $G^{(1)}_B ( 1 , 1 )$. Similarly, one can prove that

\begin{equation} \label{Cross}
\begin{array}{ccl}
\langle \mathcal{E}^{(-)}_B ( 1 ) \mathcal{E}^{(+)}_A ( 1 ) \rangle & = & \langle \mathcal{E}^{(-)}_A ( 1 ) \mathcal{E}^{(+)}_B ( 1 ) \rangle^{*} \\
& = & 4 ( \frac{\omega_0}{c} )^4 \frac{1}{r_{1}^{2}} \ \mid \rho_{12} \mid^2 \ ( \frac{\mathcal{D}^2}{6} ) \ \mid \hat z \cdot \vec \epsilon_1 \mid^2 \ e^{- i k (\vec n_1 - \vec n_l) \cdot ( \vec R_A - \vec R_B )}
\end{array}
\end{equation}

from which the complete expression for the far field intensity distribution can be derived (see (\ref{intensity}) and (\ref{efieldproj1}))

\begin{equation} \label{intense}
\begin{array}{ccl}
\langle I ( 1 ) \rangle & = & \ \langle ( \mathcal{E}^{(-)}_A ( 1 ) + \mathcal{E}^{(-)}_B ( 1 ) ) \ ( \mathcal{E}^{(+)}_A ( 1 ) + \mathcal{E}^{(+)}_B ( 1 ) ) \ \rangle\\
& = & ( G^{(1)}_A ( 1 , 1 ) + G^{(1)}_B ( 1 , 1 ) + \langle \mathcal{E}^{(-)}_B ( 1 ) \mathcal{E}^{(+)}_A ( 1 ) \rangle +  \langle \mathcal{E}^{(-)}_A ( 1 ) \mathcal{E}^{(+)}_B ( 1 ) \rangle ) \\
& = & \frac{1}{r_{1}^{2}} ( \frac{\omega_0}{c} )^4 \ ( \frac{\mathcal{D}^2}{3} ) \ ( 2 \rho_{11} ) \lbrack 1 + \frac{\mid 2 \rho_{12} \mid^2}{2 \rho_{11}} \mid \hat z \cdot \vec \epsilon_1 \mid^2 \cos ( k (\vec n_1 - \vec n_l) \cdot ( \vec R_A - \vec R_B )) \rbrack \ ,
\end{array}
\end{equation}

with, according to (\ref{masterss2})

\begin{equation} \label{fract}
\frac{\mid 2 \rho_{12} \mid^2}{2 \rho_{11}} \mid \hat z \cdot \vec \epsilon_1 \mid^2 = \frac{( \gamma_0 + \gamma )^2}{( 2 g^2 + ( \gamma_0 + \gamma )^2 )} \ \mid \hat z \cdot \vec \epsilon_1 \mid^2
\end{equation}

As can be seen from (\ref{intense}), the depth of modulation of the far field intensity distribution is determined by (\ref{fract}). Note that this factor not only goes to zero for $\sigma$-polarized fluorescence light (via the term $\mid \hat z \cdot \vec \epsilon_1 \mid$) but also in the limit of strong driving fields \cite{walls97,polder76}. The later is also true for two coherently driven 2-level atoms \cite{skornia,beige,richter91}. Here one finds for the far field intensity distribution \cite{skornia}

\begin{equation} \label{intense111}
\langle I ( 1 ) \rangle \sim 
\frac{2 g^2}{( 2 g^2 + \gamma^2 )} \left[ 1 + \frac{\gamma^2}{( 2 g^2 + \gamma^2 )} \cos ( k ( \vec n_1 - \vec n_l) \cdot ( \vec R_A - \vec R_B ) ) \right]
\end{equation}

Again, the depth of modulation $ \gamma^2 / ( 2 g^2 + \gamma^2 )$ vanishes in case of increasingly high laser power.

\vspace{1cm}

{\bf Interferences in intensity-intensity correlations}

\vspace{0.5cm}

The question thus remains: is there a possibility of exhibiting quantum interferences even in case of $\sigma$-polarized emitted light or in the case of strong driving fields. According to our general discussion in Section \ref{g2} the answer is yes if we study intensity-intensity correlations. For this purpose we turn our attention to the analysis of the crucial object (\ref{g1}). On using (\ref{g1}) and (\ref{Xi1}) we find:

\begin{equation} \label{G12}
\begin{array}{ccl}
 G^{(1)}_A ( 1 , 2 ) & = & ( \frac{\omega_0}{c} )^4 \frac{e^{i k (r_2 - r_1 )}}{r_2 r_1} \ e^{- i k ( \vec n_2 - \vec n_1 ) \cdot \vec R_A}\\
& & \lbrace ( \vec \epsilon_1^{\ast} \cdot \vec d_{21}^{\ast} ) \ ( \vec \epsilon_2 \cdot \vec d_{21} ) + ( \vec \epsilon_1^{\ast} \cdot \vec d_{43}^{\ast} ) \ ( \vec \epsilon_2 \cdot \vec d_{43} ) +\\
& & ( \vec \epsilon_1^{\ast} \cdot \vec d_{41}^{\ast} ) \ ( \vec \epsilon_2 \cdot \vec d_{41} ) + ( \vec \epsilon_1^{\ast} \cdot \vec d_{23}^{\ast} ) \ ( \vec \epsilon_2 \cdot \vec d_{23} ) \rbrace \ \cdot \rho_{11}\\
\end{array}
\end{equation}

In deriving (\ref{G12}) we used the steady state solutions (\ref{masterss1}) and (\ref{masterss2}) of the master equation for the atomic system. Note that $G^{(1)}_A ( 1 , 2 )$ is proportional to the population in the upper state which is nonzero as long as the system is excited. The term in the curly bracket in (\ref{G12}) can be shown to be proportional to $( \vec \epsilon_1^{\ast} \cdot \vec \epsilon_2 )$. Using (\ref{G12}) in (\ref{Gamma}) we thus obtain a very simple  result:

\begin{equation} \label{Gamma1}
\Gamma^{(2)} ( 1 , 2 ) = \mid \vec \epsilon_1^{\ast} \cdot \vec \epsilon_2 \mid^2 \cdot \cos \lbrace k \lbrack ( \vec n_1 - \vec n_2 ) \cdot ( \vec R_A - \vec R_B ) \rbrack \rbrace
\end{equation}

Clearly $\Gamma^{(2)} = 0$ if one decides to pick up orthogonal polarizations at the two detectors. For identical polarization we have $\vec \epsilon_1^{\ast} \cdot  \vec \epsilon_2 = 1$ so that one obtains 100\% modulation depth in the intensity-intensity correlation even if $\vec \epsilon$ corresponds to $\sigma$-polarized radiation. According to (\ref{Gamma1}) the depth of modulation $M$ is simply determined by the two plarization vectors

\begin{equation} \label{mod}
M = \mid \vec \epsilon_1^{\ast} \cdot \vec \epsilon_2 \mid^2
\end{equation}

Note that this factor obviously does not depend on the strength of the driving field exciting the system!

Finally, in order to understand the nonclassical nature of the emitted light it is useful to introduce the normalized intensity-intensity correlation function $g^{(2)} ( 1 , 2 )$ via \cite{agarwal74}

\begin{equation} \label{gnorm}
g^{(2)} ( 1 , 2 ) = \frac{ G^{(2)} ( 1 , 2 )}{(I(1) I(2))} \ ,
\end{equation}

which on using Eqs. (\ref{intense}) and (\ref{Gamma1}) simply becomes

\begin{equation} \label{gnorm1}
g^{(2)} ( 1 , 2 ) = \frac{D^{-1} (1) D^{-1} (2)}{2} ( 1 + \mid \vec \epsilon_1^{\ast} \cdot \vec \epsilon_2 \mid^2 \cos \lbrace k \lbrack ( \vec n_1 - \vec n_2 ) \cdot ( \vec R_A - \vec R_B ) \rbrack \rbrace ) \ ,
\end{equation}

with 

\begin{equation} \label{di}
D ( i ) = 1 +  \frac{\mid 2 \rho_{12} \mid^2}{2 \rho_{11}} \mid \hat z \cdot \vec \epsilon_i \mid^2 \cos \lbrace k \lbrack ( \vec n_1 - \vec n_2 ) \cdot ( \vec R_A - \vec R_B ) \rbrack \rbrace \quad , \quad i = 1,2
\end{equation}

For detection of polarization such that $\hat z \cdot \vec \epsilon_1 = 0$ we obtain

\begin{equation} \label{gnorm2}
g^{(2)} ( 1 , 2 ) = \frac{1}{2} ( 1 + \mid \vec \epsilon_1^{\ast} \cdot \vec \epsilon_2 \mid^2 \cos \lbrace k \lbrack ( \vec n_1 - \vec n_2 ) \cdot ( \vec R_A - \vec R_B ) \rbrack \rbrace ) \ ,
\end{equation}

so that for measurements with a single detector one gets 

\begin{equation} \label{gnorm21}
g^{(2)} ( 1 , 1 ) = 1
\end{equation}

We recall from \cite{skornia} that the nonclassical nature of the radiated field is reflected by the violation of the inequality

\begin{equation} \label{ineqclass}  
\prod_{i=1}^{2}( g^{(2)}( i; i ) - 1 ) \geq ( g^{(2)} ( 1 , 2 ) - 1 )^2
\end{equation}

Clearly, with $g^{(2)} ( 1 , 2 )$ given by (\ref{gnorm1}) (or in the special case by (\ref{gnorm2})), a strong violation of this inequality is possible, indicating that the emitted light has highly nonclassical properties.

\section{Conclusions} \label{concl}

In conclusion we have demonstrated interference phenomena in the second order correlations of the fluorescence light from ion pairs. The phenomena occur even in situations where first order interferences do not appear. An explicit manifestation of this phenomenon is presented by analyzing the system of two 4-level ions driven by a linearly polarized coherent laser field. In this case the far field intensity distribution of the scattered $\sigma$-polarised light does not exhibit first order interferences. By contrast, well defined interferences are displayed in the second order correlations of the emitted fluorescence light. Here we find the rather remarkable result that the intensity-intensity correlations can show a modulation depth of up to 100\% for both $\sigma$ and $\pi$-polarised fluorescence light, independent of the strength of the driving laser. In contrast, the modulation of the intensity profile, whenever present, depends strongly on the power of the driving laser field, vanishing in the limit of very high laser intensities. Finally we discussed the nonclassical nature of the emitted fluorescence radiation. As can be seen from Eqs. (41) - (45), the nonclassical nature of the scattered light can be displayed independent of the polarization of the emitted light or the strength of the driving field, i.e. independent of the appearance or disappearance of first order interference fringes. The observed interferences in the intensity-intensity correlations can also be explained in terms of the entanglement induced by the detection of the first photon. Needless to say that our results can be generalised to include more atoms or ions and that more complex atomic transitions can be considered. The former should be especially relevant to experiments with ion chains or with atoms trapped in an optical lattice.

\section{Acknowledgement} 

J. von Zanthier and G. S. Agarwal thank A. Beige for discussions on the experiment of Eichmann et al.

\vspace{3cm}

{\large \textbf{Figure Captions}}

\vspace{1cm}

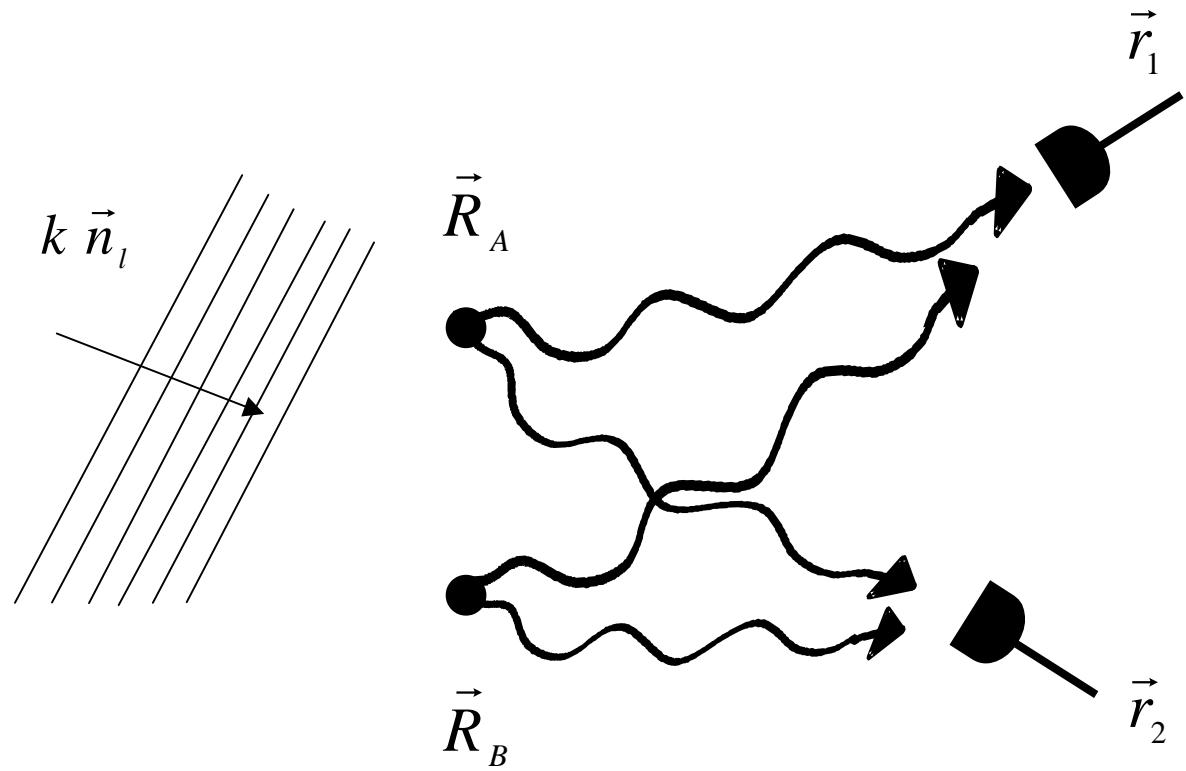
{\bf Fig. 1}: Two atom system considered: a plane wave with wave vector $k \vec n_l$ is impinging on two atoms fixed at positions $\vec R_A$ and $\vec R_B$. The light scattered by the two atoms is registered in the far field by two detectors positioned at $\vec r_1$ and $\vec r_2$.

\vspace{1cm}

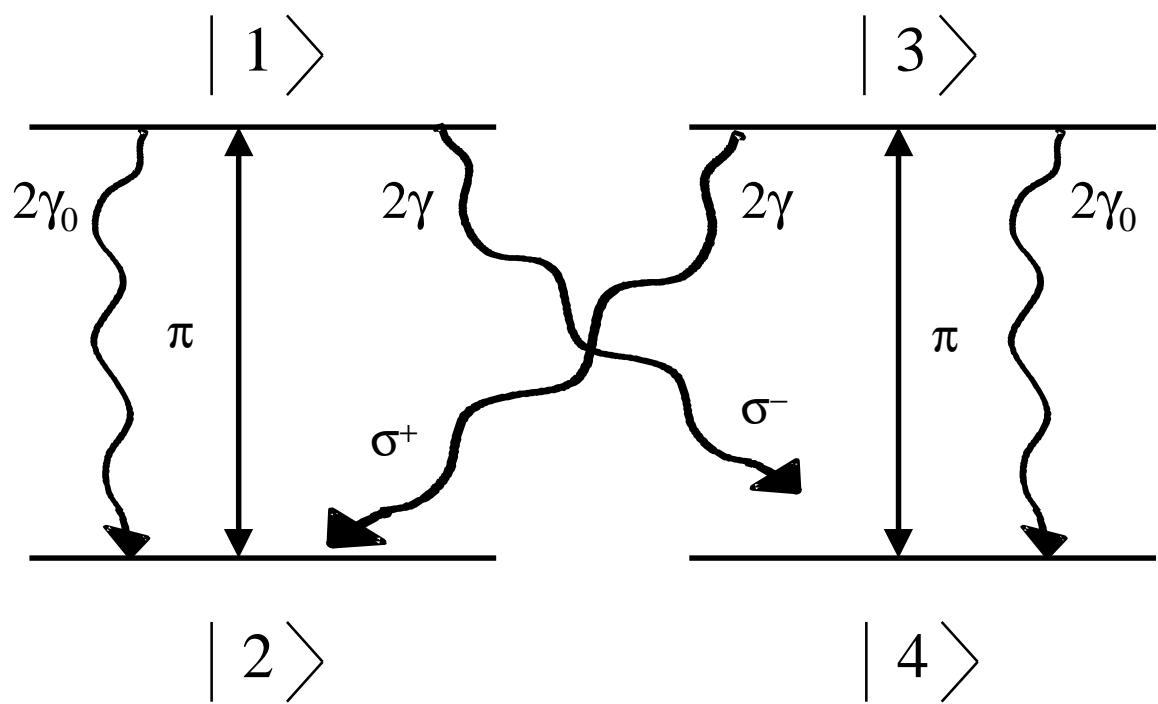
{\bf Fig. 2}: $ J = 1/2 \rightarrow J = 1/2$ internal level scheme of the two atoms. Both atoms are excited by $\pi$-polarized light.

\end{document}